\documentclass[12pt,preprint]{aastex}

\shorttitle{Christodoulou et al.}
\shortauthors{Interchange Method in Compressible Couette Flow}

\begin{document}


\title{Interchange Method in Compressible Magnetized Couette Flow: \\
	 Magnetorotational and Magnetoconvective Instabilities}


\author{Dimitris M. Christodoulou,\altaffilmark{1} John Contopoulos,\altaffilmark{2}
	  and Demosthenes Kazanas\altaffilmark{3}}


\altaffiltext{1}{Lucent Technologies, 1 Robbins Road, Westford, MA 01886.
			E-mail: dmc12@lucent.com}
\altaffiltext{2}{ELVAL, 200 Akti Themistokleous Street, Piraeus 18539, Greece.
			E-mail: jcontop@elval.vionet.gr}
\altaffiltext{3}{NASA/GSFC, Code 665, Greenbelt, MD 20771.
			E-mail: kazanas@milkyway.gsfc.nasa.gov}


\baselineskip = 12pt

\begin{abstract}

\def\sles{\lower2pt\hbox{$\buildrel {\scriptstyle <}
   \over {\scriptstyle\sim}$}}
\def\sgreat{\lower2pt\hbox{$\buildrel {\scriptstyle >}
   \over {\scriptstyle\sim}$}}

We obtain the general forms of the axisymmetric stability criteria in
a magnetized compressible Couette flow using an energy variational
principle, the so--called interchange or Chandrasekhar's method, which
we applied successfully in the incompressible case.  This formulation
accounts for the simultaneous presence of gravity, rotation, a
toroidal magnetic field, a weak axial magnetic field, entropy gradients, 
and density gradients in the initial equilibrium state.  The power of the
method lies in its simplicity which allows us to derive extremely
compact and physically clear expressions for the relevant stability
criteria despite the inclusion of so many physical effects.  In the
implementation of the method, all the applicable conservation laws are
explicitly taken into account during the variations of a quantity with
dimensions of energy which we call the ``free--energy function."  

As in the incompressible case, the presence of an axial
field invalidates the conservation laws of angular momentum and azimuthal
magnetic flux and introduces instead isorotation and axial current
conservation along field lines. Our results are therefore markedly different
depending on whether an axial magnetic field is present, and generalize 
in two simple expressions all previously known, partial stability criteria for 
the appearance of magnetorotational instability.
Furthermore, the coupling between magnetic tension and buoyancy and its
influence to the dynamics of nonhomoentropic magnetized flows becomes 
quite clear from our results. In the limits of plane--parallel atmospheres
and homoentropic flows, our formulation easily recovers the stability 
criteria for suppression of convective and Parker instabilities, as well as 
some related special cases studied over 40 years ago by Newcomb and Tserkovnikov 
via laborious variational techniques.

\end{abstract}





\section{Introduction}

\subsection{Motivation}

There has been a recent controversy about the direction of transport 
of angular momentumin in radiatively inefficient accretion disk theory,
the theory most successful in addressing the dynamics and radiation 
emission in low--luminosity accretion flows. Two schools of
thought have come to different conclusions by analyzing two
opposite limiting cases: Proponents of magnetorotational
instability \citep{bh2,bh3,ha} largely ignore convection and find
generic outward transport of angular momentum,
while proponents of convective instability largely ignore magnetic fields
and find on average slow inward transport of angular momentum
by the combination of viscous forces and buoyancy \citep{ia1,ia2,spb,
ian,nia,qg,aiqn}. 

It is our opinion that this controversy stems from our ignorance 
of the precise stability criterion of nonhomoentropic, magnetized, 
differentially rotating flows that are thought to be associated
with the aforementioned low--efficiency accretion onto black holes.
Such a criterion has eluded 
standard linearization and energy variational techniques because the
corresponding set of perturbed axisymmetric equations is quite complex 
and the exact problem has never been reduced to a single modal equation. 
For this reason and in view of its current importance to the 
understanding of low--luminosity accretion onto black holes in X-ray 
binaries and active galactic nuclei (see Narayan [2002] for a review), 
we revisit this problem employing the powerful 
free--energy variational method developed by \citet{ckst1,ckst2} 
and applied to incompressible Couette flows 
in an earlier publication (Christodoulou et al. 1996, hereafter Paper~I).

Very recently, \citet{bh3} and \citet{nqia} attempted to tackle the same problem
with a set of linearized equations derived by \citet{bh1} by augmenting
them with an additional radial convective term. However, it has been
pointed out by \citet{knob} and re-iterated in Paper~I that the 
original set of equations does not take into account the effect of a 
curvature--dependent magnetic--tension term. But it is precisely
this term that couples into the conventional 
Brunt-V$\ddot{\rm a}$is$\ddot{\rm a}$l$\ddot{\rm a}$ frequency and modifies
the effect of buoyancy in magnetized nonhomoentropic flows. As we shall see
below, the influence of the magnetic tension is particularly dramatic in
flows with combined toroidal and poloidal magnetic fields, where it can
single-handedly change the sign of the convective term in the stability
criterion, thereby reversing the effect of superadiabatic and subadiabatic
entropy gradients relative to what is known from the conventional
Schwarzschild criterion.

While the results of our study shed some light to the above--mentioned 
controversy, we believe that they are also of considerably broader interest.
We expect the results described in this paper to prove useful in
undestanding both the global structure of such accretion flows and the 
associated jet--like outflows; as well as in other areas of astrophysics
and likely in branches of plasma/fusion physics which involve the 
combined effects of convection, differential rotation, and magnetic fields.
This generality is not surprising when considering that some of the hydromagnetic 
stability criteria discussed below have been previously derived in studies of laboratory 
plasmas and planetary magnetospheres (see, e.g., Newcomb [1962] 
and Rogers \& Sonnerup [1986], respectively).

\subsection{Previous Work}

In Paper~I we analyzed the stability of magnetized
incompressible Couette flows using an energy variational principle,
the so--called interchange or Chandrasekhar's (1960) method.  This
formulation reproduced previously known stability criteria in such an
elegant and simple manner, that we became convinced that the
interchange method is indeed physically correct, mathematically
robust, efficient, and an ideal technique for analyzing a variety of 
difficult stability problems. Motivated by this success we proceed in
this work to extend the interchange method to the stability 
of axisymmetric {\em compressible nonhomoentropic} magnetized Couette
flows.  In doing so, it becomes necessary to clarify a number of
crucial details concerning the application of the method: Why do the
"isotropic" pressure forces not contribute to the construction of
the free--energy function? What is the role of gravity during the
simultaneous interchange of two fluid elements?  What are the rules
according to which the interchange of compressible fluid elements 
is performed? Does the interchange involve elements of equal mass or
equal volume?

Understanding the answers to these questions sheds light to the
physics of the studied instabilities as well as to the physical
meaning of the interchange method itself. For example, it becomes
clear why we should simultaneously interchange two fluid elements and
not simply displace/perturb one element from its original position.
Such a single--displacement approach was discussed 
by \citet{rs} for the case of uniform rotation in an homoentropic 
fluid and led to results which did not allow one to discern the basic
physics of the stability problem. Rogers \& Sonnerup also used the
standard interchange method of two fluid elements, but they over-constrained
the volumes of the fluid elements attempting to interchange equal masses.
Their result is valid for uniform rotation, but it is doubtful whether
the same methodology can be productive in more complex (differentially
rotating and nonhomoentropic) fluids.

Interestingly enough, \citet{rs} were trying to understand 
why a calculation of the
stability of magnetized compressible flows previously performed by
\citet{cheng} had led to two different stability criteria depending on
the use of a purely convective stability criterion or the interchange method.
Cheng argued that the interchange method produces incorrect
results.  Unfortunately, the disagreement is entirely due to
the incorrect implementation of the interchange method in Cheng's
investigation. In particular, Cheng demanded that the interchanged fluid
elements be in pressure balance after the interchange. This condition is
not appropriate which explains why the interchange method produced the wrong
result. The inconsistency in assuming pressure balance after the interchange
is easily identified in the ``free--energy" variational framework defined by
\citet{ckst1,ckst2}: When fluid elements are interchanged, the system is effectively
taken out of equilibrium and the only relevant question is whether the free energy
increases or decreases. As was graphically and analytically shown in \citet{ckst1}, 
a decrease in the out-of-equilibrium free energy indicates the presence of another
equilibrium state of lower energy and the system will evolve ``downhill" in energy
space seeking
this new equilibrium state. Whether the system will get there on a dynamical
timescale (or in an Alfv\'en time as Cheng remarked) depends purely on the
applicable conservation laws, but this is of no consequence to 
the variational method and certainly does not imply that the anticipated
pressure balance in the final equilibrium state should be built into
the interchange method. 

\citet{new2} has also analyzed the same problem in the special case
of a non-gravitating homoentropic fluid by using a Hamiltonian formulation along with
an energy variational principle and was able to recognize the importance of the
conservation law of angular momentum in the case of a Couette flow
threaded only by a toroidal magnetic field and the importance of
isorotation in the case of a Couette flow threaded only by a weak
axial field.  Unfortunately, Newcomb's method is rather complicated
and mathematically tedious, a fact that quickly obscures the physics 
of the problem. These difficulties have prevented \citet{new2} from 
obtaining a stability criterion in the combined case of both toroidal 
and weak axial magnetic--field components.  We discuss in more
detail the above-mentioned older investigations in an Appendix to this paper, 
where we analyze various special cases of our general results.

\subsection{Outline}

In the following sections, we describe the results that we have
obtained on the above stability problems by our free--energy
variational method and we clarify the role of various force components
and conservation laws in the general case of rotating, gravitating,
magnetized, compressible, nonhomoentropic flows.  In \S~2, we discuss
details of the implementation of the interchange method to the
axisymmetric stability of compressible flows. In \S~3, we analyze the
case of Couette flow threaded by a purely toroidal magnetic field.  In
\S~4, we demonstrate that the introduction of a weak axial--field component to
the model of \S~3 effects formidable changes to the applicable
conservation laws and to the resulting stability characteristics of
the flow.  In \S~5, we apply our method to static plane--parallel
atmospheres under the influence of gravity and we recover well--known
criteria (Schwarzschild 1906; Tserkovnikov 1960; Newcomb 1961; Parker
1966, 1975, 1979) for suppression of convective motions and magnetic
buoyancy.  Finally, in \S~6, we discuss astrophysical applications of
our results in the context of accretion flows and magnetically--driven
jet outflows. In an Appendix to the paper, we reduce our results
by selectively ignoring some of the physics involved (entropy gradients
and either gravity or differential rotation), and we recover 
all previously known partial stability criteria \citep{ts,new2,rs}.

\section{The Interchange Method}

Let us first describe the interchange method for compressible flows.
We adopt cylindrical coordinates ($r, \phi, z$) and consider
axisymmetric perturbations applied to axisymmetric
($\partial/\partial\phi =0$) magnetized Couette flows.  The magnetic
field is assumed to be frozen--in to the fluid.  Following Rayleigh
(see, e.g., Chandrasekhar 1981, \S~66), we consider the simultaneous radial
interchange of two nearby fluid elements orbiting
initially at radii $r_1$ and $r_2>r_1$. In order to perform the
interchange correctly in the general compressible case, we need to
account for the following important details:
\begin{enumerate}
\item The initial configuration is in hydromagnetic equilibrium.
\item The interchange involves two fluid elements of equal volume, not
equal mass. During and at the end of the interchange, the ambient
fluid must retain its continuity and the method should not perturb or modify
any boundaries that may exist in the fluid. Therefore, we should
make sure that each displaced fluid element occupies exactly the
volume left vacant by the displacement of the other fluid element.
\item Volume continuity also implies that, although the sum of the 
two masses of the interchanged elements is conserved, individual masses are not.
Therefore, unequal masses are interchanged which is also necessary in the presence of
gravity, if the fluid elements are going to feel the effects of the gravitational
field.
\item The interchange is performed adiabatically, i.e., in a way
that certain physical parameters which express conservation laws
associated with each fluid element remain constant during the
interchange. Where applicable, the conservation laws must be expressed 
per unit mass since individual element masses are not conserved.
\item The free--energy variational principle is applied between
the initial equilibrium state and the new nonequilibrium end
state.  What will happen afterwards, e.g. whether the interchanged 
fluid elements will expand or contract in their new positions
and will disturb the ambient fluid locally, is of no consequence to
the method.
\end{enumerate}

It is physically important to realize that if the change in free
energy after the interchage is positive (negative), then the initial
equilibrium is dynamically stable (unstable) to axisymmetric
perturbations and the exchanged fluid elements are subject to return
to (escape from) their original equilibrium positions (cf. Christodoulou et al. [1995a,b]).
In implementing the method, we first calculate the
initial condition for equilibrium and the free energy spent or gained due to
the action of radial forces.  Since the unperturbed fluid is in
equilibrium, the total radial force per unit mass $F (r)$ at every
point of the flow is zero, viz.
\begin{equation}
F(r)\equiv \frac{v_\phi^2}{r}-\frac{B_\phi^2}{\rho r}
-\frac{1}{\rho}\frac{d}{d r}
\left(P_{fl}+\frac{B_\phi^2+B_z^2}{2}\right)
+ g_r = 0\ ,
\label{eq}
\end{equation}
where $v_\phi$ is the azimuthal flow velocity; $B_\phi$ and $B_z$ are
the toroidal and axial components of the magnetic field in such units
that the magnetic permeability is $\mu = 4\pi$; $\rho$ and
$P_{fl}$ are the density and internal nonmagnetic pressure of the ideal fluid, 
respectively; and $g_r\equiv
-{d\Phi_{\rm grav}}/{d r}$ is the radial force per unit mass
due to the overall gravitational potential $\Phi_{\rm grav}$.
Eq.~(\ref{eq}) accounts for all the forces which act radially on each
fluid element in equilibrium: the centrifugal force, the tension of
the magnetic field, the fluid and magnetic--field pressure gradients,
and the gravitational force. 

The change in free energy $\Delta E$ of the configuration is defined
to be equal to the work done in performing the interchange between
the positions $r_1$ and $r_2$, viz.
\begin{equation}
\Delta E \equiv -\int_{1\rightarrow 2} m F(r)dr-\int_{2\rightarrow 1} m F(r)dr
-\int_{1\rightarrow 2} m P dV
-\int_{2\rightarrow 1} m P dV\ ,
\label{DE}
\end{equation}
where the notation $i\rightarrow j$ with $i,j=1,2$ is used to indicate
that the integral is taken during the displacement of fluid element
$i$ from position $i$ to position $j$; $m=\rho V$ is the element mass; 
the change in volume
$dV\equiv 0$ in this method, as shown explicitly below; and $P$ denotes the 
{\em total} internal pressure of each displaced fluid element, i.e.
\begin{equation}
P\equiv P_{fl} + P_{mag} = K \rho^\gamma + \frac{B_\phi^2+B_z^2}{2}\ ,
\label{PPP}
\end{equation}
where $P_{mag}$ is the magnetic pressure, $K$ is a factor that depends
on the specific entropy of the fluid, and $\gamma$ is the adiabatic
index.  $K$ is constant for homoentropic fluids and remains the same
during slow adiabatic interchanges, but varies in space if an entropy
gradient is present in the fluid.

Mass conservation in this method is expressed by the requirement that the total
mass of the two fluid elements be the same before and after the interchange.
Suppose that the interchange involves two fluid elements initially at radii
$r_1$ and $r_2>r_1$ with masses $m_1=\rho_1 V_1$ and $m_2=\rho_2 V_2$,
respectively. After the interchange, the two masses are $m_1^\prime
=\rho_1^\prime V_2$ and $m_2^\prime =\rho_2^\prime V_1$ at radii
$r_2$ and $r_1$, respectively. Since the densities change
adiabatically, i.e., $\rho_1^\prime = \rho_1 +\Delta\rho_{ad}$ and
$\rho_2^\prime = \rho_2 -\Delta\rho_{ad}$ to linear order, then the
conservation of total mass, $m_1+m_2=m_1^\prime +m_2^\prime$, leads to
the constraint $(\Delta\rho-\Delta\rho_{ad}) (V_2-V_1) = 0$, 
where $\Delta\rho\equiv\rho_2 - \rho_1$.  Since,
in general, $\Delta\rho\neq\Delta\rho_{ad}$, equal volumes must be
adopted for adiabatic interchanges of individual fluid elements.

It is now easy to understand the physics of the interchange method: During
and after the interchange of two fluid elements, the internal pressure is not a driving
force; it simply continues to adjust in trying to maintain the continuity of the
fluid and the conserved quantities. Therefore, the pressure--gradient term in eq.~(\ref{eq})
does not contribute to the integrals of eq.~(\ref{DE}). On the other hand,
the plus or minus signs of the purely radial forces are not changed by the interchange
and these forces
maintain the same unidirectional behavior. Only their magnitudes are changed when
these terms are properly constrained by the applicable conservation laws during
the displacements from radius $r_i$ to $r_j$.
Therefore, the radial forces will continue to drive the interchanged fluid
elements either back toward 
their initial equilibrium positions (restoring forces) or away from them (instability forces)
and their combined contribution (in both magnitude and direction) to the free energy 
decides whether the system will evolve uphill or downhill in energy space. Of course,
in the former case the system is bound to return back to the initial state of lower
energy, while in the latter case it runs away from the initial equilibrium and toward
nonequilibrium states of lower total energy.

Since $V_1=V_2\equiv V$ in the interchange method, we are then left with 
the following expression for the change in free energy written now per unit volume:
\begin{equation}
\frac{\Delta E}{V} = \int_{1\rightarrow 2}\left(
-\frac{\rho v_\phi^2}{r}+\frac{B_\phi^2}{r} - \rho g_r\right)dr
+\int_{2\rightarrow 1}\left(
-\frac{\rho v_\phi^2}{r}+\frac{B_\phi^2}{r} - \rho g_r\right)dr .
\label{DE2}
\end{equation}
As was mentioned above, a necessary and sufficient condition for stability 
to interchange of equal--volume elements is that
\begin{equation}
\frac{\Delta E}{V} \geq 0\ .
\label{condition}
\end{equation}
In the incompressible case, the fluid density $\rho$ is constant
and the results of Paper~I are
obtained by simply evaluating the two integrals after the
relevant conservation laws have been built into the force terms
of each integrand.  We now proceed to obtain the relevant
stability criteria for compressible fluids in the two fundamentally
different (due to differences in the conserved quantities) cases of
interest (cf. Paper~I): a Couette flow with a purely toroidal magnetic
field, and a Couette flow with both toroidal and axial field
components.

\section{Purely Toroidal Magnetic Fields}

As was already strongly emphasized in Paper~I, before we can calculate the
above integrals, we must take into account all the physical quantities
which are conserved during the interchange. Clearly, total mass is
conserved during the interchange (\S~2). Furthermore, because only a toroidal
magnetic field $B_\phi (r)$ is assumed to be present, the specific
(i.e. per unit mass) angular momentum and azimuthal magnetic flux are
also conserved during the interchange (see also Paper~I).\footnote{We
note that, under the axisymmetric conditions assumed throughout \S~3
and \S~4, specific angular momentum is identical to circulation.  For
nonaxisymmetric perturbations, however, it is the circulation, and not
the specific angular momentum, which dictates a conservation law.}
Finally, we assume that each fluid element conserves its specific
entropy during interchanges.  
Note that, although the variations of the free energy are expressed
per unit volume for convenience, the conserved quantities must be expressed 
per unit mass
because individual element masses are not conserved during interchanges.
Then, the above conservation laws are
expressed mathematically by the following set of equations:
\begin{eqnarray}
   m_i + m_j & = & {\rm const} , \\
   L_i \equiv \left[r v_\phi\right]_i & = & {\rm const} , \\
   \Phi_i \equiv \left[\frac{B_\phi}{\rho r}\right]_i & = & {\rm const} , \\
   K_i \equiv \left[\frac{P_{fl}}{\rho^\gamma}\right]_i & = & {\rm const} , \
\end{eqnarray}
where $m_i$ and $m_j$ are the initial masses of the two elements; and the symbols
$L_i$, $\Phi_i$, and $K_i$ refer to the conserved quantities in each
fluid element $i$ (specific angular momentum, specific azimuthal
magnetic flux, and specific entropy, respectively).  At this point, it
should be noted that, owing to the various gradients present in the fluid,
the values of $L_i$, $\Phi_i$, and $K_i$ {\em
are generally different} between any two fluid elements located at two
different radii $r_i$ and $r_j$ in the equilibrium flow.

Before we go on to obtain the stability criterion, we define
what we mean by ``adiabatic gradients" in fluids threaded by a toroidal
field only. An arbitrary perturbation in density $d\rho$ leads to an 
adiabatic change in pressure $dP_{ad}$ if the change occurs under
the constraints imposed by the applicable conservation laws
($K$=const and $\Phi$=const). In our case
\begin{equation}
P = K\rho^\gamma+\frac{1}{2}\Phi^2 \rho^2 r^2\ ,
\label{pres_phi}
\end{equation}
and
\begin{equation}
\left.\frac{dP}{dr}\right|_{ad} = c_\phi^2\frac{d\rho}{dr} + \frac{B_\phi^2}{r}\ .
\label{dpres_ad_phi}
\end{equation}
The last term is written as $B_\phi^2/r$ instead of $r\rho^2\Phi^2$ for convenience. 
The symbol $c_\phi$ represents the fast magnetosonic speed in the $\phi$ direction:
\begin{equation}
c_\phi^2 \equiv \frac{\gamma P_{fl} + B_\phi^2}{\rho} \ .
\label{c_phi}
\end{equation}
The above definitions generalize the well--known hydrodynamic expression
$dP_{fl}/dr|_{ad}=c_o^2 d\rho /dr$ in which the adiabatic sound speed $c_o$ in
the fluid is given by $c_o^2=\gamma P_{fl}/\rho$.  

In a similar fashion, an arbitrary perturbation in pressure $dP$ leads to an 
adiabatic change in density $d\rho_{ad}$ if the change occurs under
the constraints imposed by the applicable conservation laws
($K$=const and $\Phi$=const). Then we find that
\begin{equation}
\frac{dP}{dr} = c_\phi^2\left.\frac{d\rho}{dr}\right|_{ad} + \frac{B_\phi^2}{r}\ .
\label{dpres_phi}
\end{equation}
Combining now eqs.~(\ref{dpres_ad_phi}) and (\ref{dpres_phi}), we find that
\begin{equation}
\frac{dP}{dr} - \left.\frac{dP}{dr}\right|_{ad} = 
c_\phi^2\left(\left.\frac{d\rho}{dr}\right|_{ad} - \frac{d\rho}{dr}\right) \ .
\label{ad}
\end{equation}
The magnetic field enters this equation implicitly 
through $c_\phi$ and the pressure terms.
We shall see in \S~4 that this behavior does not generalize in the presence
of both $B_\phi$ and $B_z$ components, thus eq.~(\ref{ad}) is not generally
applicable to all magnetic-field configurations.
 
We are now ready to calculate and reduce the integrals shown in eq.~(\ref{DE2}).
Using the trapezoidal rule to evaluate each of the interchanges $1\to 2$ and $2\to 1$, 
we find that
\begin{eqnarray}
& \frac{\Delta E}{V} = \nonumber\\
& \frac{1}{2}\left[ 
- \left\{ \frac{\rho_1 L_1^2}{r_1^3}
+ \frac{(\rho_1 +\Delta\rho_{ad}) L_1^2}{r_2^3} \right\}
+ \left\{ r_1 \rho_1^2 \Phi_1^2 + r_2 (\rho_1 +\Delta\rho_{ad})^2 \Phi_1^2 \right\} 
- \left\{ g_{r1} \rho_1 + g_{r2} (\rho_1 +\Delta\rho_{ad}) \right\} 
\right]\left(+\Delta r\right) \nonumber\\
+ & \frac{1}{2}\left[ 
- \left\{ \frac{\rho_2 L_2^2}{r_2^3}
+ \frac{(\rho_2 -\Delta\rho_{ad}) L_2^2}{r_1^3} \right\}
+ \left\{ r_2 \rho_2^2 \Phi_2^2 + r_1 (\rho_2 -\Delta\rho_{ad})^2 \Phi_2^2 \right\} 
- \left\{ g_{r2} \rho_2 + g_{r1} (\rho_2 -\Delta\rho_{ad}) \right\}
\right]\left(-\Delta r\right) . \nonumber\\
\label{big1}
\end{eqnarray}
Keeping terms up to second order in the expansions of eq.~(\ref{big1}), 
eqs.~(\ref{condition}) and (\ref{big1}) give
\begin{equation}
\frac{\Delta E}{V} = \rho\Delta r\left[\frac{\Delta L^2}{r^3}
-\rho r\Delta\Phi^2 
+ \left(\frac{L^2}{r^3}-2r\rho\Phi^2+g_r\right)
\frac{\Delta \rho-\Delta \rho_{ad}}{\rho} \right] \geq 0\ ,
\label{dedr}
\end{equation}
where $\Delta$-quantities represent centered "2 minus 1" differences 
(e.g., $\Delta L^2 = L_2^2 - L_1^2$) and the remaining 
quantities represent averages over positions $r_1$ and $r_2>r_1$
(e.g., $m=(m_1 + m_2)/2$).
Using eqs.~(\ref{eq}) and (\ref{ad}), eq.~(\ref{dedr}) becomes equivalent 
to the necessary and sufficient criterion for stability to axisymmetric
perturbations
\begin{equation}
\frac{1}{r^3}\frac{dL^2}{d r}-\rho r\frac{d\Phi^2}{d r}
-\frac{1}{\rho^2 c_\phi^2}\left(\frac{dP}{dr}-\frac{B_\phi^2}{r}\right)
\left(\frac{dP}{dr} - \left.\frac{dP}{dr}\right|_{ad} \right) \geq 0\ ,
\label{criterion1}
\end{equation}
with $L\equiv r v_\phi$ and $\Phi\equiv B_\phi /(\rho r)$.  

Eq.~(\ref{criterion1}) constitutes the first
important result of this work.  It is interesting that, apart from the
new magnetoconvective term, eq.~(\ref{criterion1}) has basically the
same structure as the stability criterion derived for the much simpler
incompressible case (see eq.~[3.5] in Paper~I where the constant
$\rho$ was normalized away). Of course, the above stability
criterion also involves implicitly derivatives of the equilibrium
density, a quantity that plays no role in the incompressible case. 

The new magnetoconvective term in eq.~(\ref{criterion1}) reveals the
coupling between the magnetic tension and the pressure differential
that drives buoyancy: In the absence of $B_\phi$ field, this
term simply introduces the hydrodynamic
Brunt-V$\ddot{\rm a}$is$\ddot{\rm a}$l$\ddot{\rm a}$ 
frequency $N_o$ given by the expression (Tassoul 1978)
\begin{equation}
N_o^2 \equiv \frac{1}{\rho}\frac{dP_{fl}}{dr}
\left( \frac{1}{\rho}\frac{d\rho}{dr} - 
\frac{1}{\gamma P_{fl}}\frac{dP_{fl}}{dr} \right)
= - \frac{1}{\rho^2 c_o^2}\frac{dP_{fl}}{dr}
\left(\frac{dP_{fl}}{dr} - \left.\frac{dP_{fl}}{dr}\right|_{ad}\right) 
\ ,
\label{bv}
\end{equation}
but in the presence of $B_\phi$, convective motions in
astrophysical fluids are actually strengthened by magnetic tension. 
This is because in gravitating fluids, pressure gradients are intrinsically 
negative and $B_\phi^2/r$ enters  the second coefficient of the magnetoconvective 
term also with a negative sign. The result of magnetic tension then is to increase 
the magnitude of the last term of eq.~(\ref{criterion1}) without affecting its sign.
Consequently, both the frequencies of stable, oscillatory waves and the growth rates of
unstable magnetoconvective modes (if any) are increased by the $B_\phi^2/r$ term accordingly. 
The behavior of the magnetic tension in this case is determined solely by the 
imposed conservation of specific azimuthal magnetic flux at any given radial position $r$.

Finally we note that the magnetoconvective term does not disappear in homoentropic
fluids with $dK/dr$=0 because total pressure depends implicitly on the magnetic field.
For the present case with $B_\phi$ field only, using eq.~(\ref{pres_phi}) and the 
conservation laws, 
the pressure differential of eq.~(\ref{ad}) can be written in the equivalent form
\begin{equation}
\frac{dP}{dr} - \left.\frac{dP}{dr}\right|_{ad} = 
 \rho^\gamma \frac{dK}{dr} + \frac{1}{2}\rho^2 r^2 \frac{d\Phi^2}{dr}\ .
\label{dpk}
\end{equation}
This form is used in the Appendix where we combine eqs.~(\ref{criterion1})
and (\ref{dpk}) with $dK/dr$=0 to confirm Newcomb's (1962) homoentropic 
stability criterion.

\section{Toroidal and Weak Axial Magnetic Fields}

The conservation laws of total mass and specific entropy endure the
introduction of an axial magnetic field $B_z (r)$.  On the other hand,
the axial field eliminates the conservation laws of specific angular
momentum and azimuthal magnetic flux, and replaces them with the
conservation of angular velocity and of axial current along field lines. 
A detailed description of this fundamental change was given in Paper~I. 
In addition, the axial field introduces the conservation of the specific
axial magnetic flux.  The relevant conservation laws of this case are
expressed mathematically by the following set of equations:
\begin{eqnarray}
   m_i + m_j & = & {\rm const} , \\
   I_i \equiv \left[r B_\phi\right]_i & = & {\rm const} , \\
   \Omega_i \equiv \left[\frac{v_\phi}{r}\right]_i & = & {\rm const} , \\
   H_i \equiv \left[\frac{B_z}{\rho}\right]_i & = & {\rm const} , \\
   K_i \equiv \left[\frac{P_{fl}}{\rho^\gamma}\right]_i & = & {\rm const} ,
\end{eqnarray}
where the new symbols $I_i$, $\Omega_i$, and $H_i$ refer to the new
conserved quantities in each fluid element $i$ (axial current interior 
to the radius $r$,\footnote{To be more precise,
in order for the interchange to be performed without perturbing the
background fluid and its magnetic field, screening currents must be set up
at the radial surface boundaries of the displaced fluid elements. As
these screening currents are moved along with the fluid elements, the
total axial current interior to the radius $r$ of each fluid element
is indeed conserved during the displacement.} angular velocity,
and specific axial magnetic flux, respectively).

Along the lines discussed in Paper~I, we note that an axisymmetric
perturbation in a fluid threaded by an axial field will generally
induce curvature to the axial field and will therefore introduce
curvature--dependent terms in the equations.  The interchange method
that we are implementing does not include such perturbations since
$z$--dependent terms are not accounted for during interchanges.  On
the other hand, if the axial magnetic field is sufficiently weak, its
contribution to the energy change of the configuration will be
negligible compared to the other terms.  Therefore, we conclude that
the condition for stability which we are about to derive is necessary
and sufficient only if the axial wavelength of the perturbation is
infinite, or in the limit $B_z\to 0$. In general, this condition will
be only sufficient for stability, since the curvature--dependent terms have a
stabilizing influence (i.e., they make a positive contribution in the
expression for $\Delta E$).

Having identified the relevant conservation laws in this case, we can proceed 
just as in \S~3. First we define
what we mean by ``adiabatic gradients" in fluids threaded by toroidal
and axial field components. An arbitrary perturbation in density $d\rho$ leads to an 
adiabatic change in pressure $dP_{ad}$ if the change occurs under
the constraints imposed by the applicable conservation laws
($K$=const, $I$=const, and $H$=const). In this case
\begin{equation}
P = K\rho^\gamma+\frac{1}{2}\left( \frac{I^2}{r^2} + \rho^2 H^2 \right)\ ,
\label{pres_phi2}
\end{equation}
and
\begin{equation}
\left.\frac{dP}{dr}\right|_{ad} = c_z^2\frac{d\rho}{dr} - \frac{B_\phi^2}{r}\ .
\label{dpres_ad_phi2}
\end{equation}
The last term is written as $B_\phi^2/r$ instead of $I^2/r^3$ for convenience. 
Notice the change of sign in this term relative to the corresponding term
in eq.~(\ref{dpres_ad_phi}); it is caused by the introduction of axial current
conservation in place of azimuthal magnetic flux.
The symbol $c_z$ represents the fast magnetosonic speed in the $z$ direction:
\begin{equation}
c_z^2 \equiv \frac{\gamma P_{fl} + B_z^2}{\rho} \ .
\label{c_phi2}
\end{equation}
This is the characteristic speed in this case despite the presence of
two field components. This fact alone tells us that the dynamics of the system
is now driven by the magnetic coupling of fluid elements in the $z$ direction.

In a similar fashion, an arbitrary perturbation in pressure $dP$ leads to an 
adiabatic change in density $d\rho_{ad}$ if the change occurs under
the constraints imposed by the applicable conservation laws
($K$=const, $I$=const, and $H$=const). Then we find that
\begin{equation}
\frac{dP}{dr} = c_z^2\left.\frac{d\rho}{dr}\right|_{ad} - \frac{B_\phi^2}{r}\ .
\label{dpres_phi2}
\end{equation}
Combining now eqs.~(\ref{dpres_ad_phi2}) and (\ref{dpres_phi2}), we find that
\begin{equation}
\frac{dP}{dr} - \left.\frac{dP}{dr}\right|_{ad} = 
c_z^2\left(\left.\frac{d\rho}{dr}\right|_{ad} - \frac{d\rho}{dr}\right) \ .
\label{ad2}
\end{equation}
The two field components enter this equation implicitly through the pressure terms.
$B_z$ is however the only field component that determines which magnetosonic
speed is relevant to the stability of the given configuration.

We are now ready to calculate and reduce the integrals shown in eq.~(\ref{DE2}).
Using the trapezoidal rule to evaluate each of the interchanges $1\to 2$ and $2\to 1$, 
we find that
\begin{eqnarray}
\frac{\Delta E}{V} &=& \frac{1}{2}\left[ 
- \left\{ \rho_1 r_1 \Omega_1^2 + (\rho_1 +\Delta\rho_{ad}) r_2 \Omega_1^2 \right\} 
+ \left\{  \frac{I_1^2}{r_1^3} + \frac{I_1^2}{r_2^3} \right\} 
- \left\{ g_{r1} \rho_1 + g_{r2} (\rho_1 +\Delta\rho_{ad}) \right\} 
\right]\left(+\Delta r\right) \nonumber\\
&+&\frac{1}{2}\left[ 
- \left\{ \rho_2 r_2 \Omega_2^2 + (\rho_2 -\Delta\rho_{ad}) r_1 \Omega_2^2 \right\} 
+ \left\{  \frac{I_2^2}{r_2^3} + \frac{I_2^2}{r_1^3} \right\} 
- \left\{ g_{r2} \rho_2 + g_{r1} (\rho_2 -\Delta\rho_{ad}) \right\}
\right]\left(-\Delta r\right) . \nonumber\\
\label{big2}
\end{eqnarray}
Keeping terms up to second order in the expansions of eq.~(\ref{big2}), 
eqs.~(\ref{condition}) and (\ref{big2}) give
\begin{equation}
\frac{\Delta E}{V} = \rho\Delta r\left[r\Delta\Omega^2
-\frac{\Delta I^2}{\rho r^3} 
- \left(\Omega^2 r + g_r\right)
\frac{\Delta \rho-\Delta \rho_{ad}}{\rho} \right] \geq 0\ ,
\label{dedr2}
\end{equation}
where $\Delta$-quantities represent centered "2 minus 1" differences 
(e.g., $\Delta\Omega^2 = \Omega_2^2 - \Omega_1^2$) and the remaining 
quantities represent averages over positions $r_1$ and $r_2>r_1$
(e.g., $\rho=(\rho_1 + \rho_2)/2$).
Using eqs.~(\ref{eq}) and (\ref{ad2}), eq.~(\ref{dedr2}) becomes equivalent 
to the sufficient criterion for stability to axisymmetric perturbations
\begin{equation}
r \frac{d\Omega^2}{d r} - \frac{1}{\rho r^3}\frac{dI^2}{d r}
-\frac{1}{\rho^2 c_z^2}\left(\frac{dP}{dr} + \frac{B_\phi^2}{r}\right)
\left(\frac{dP}{dr} - \left.\frac{dP}{dr}\right|_{ad} \right) \geq 0\ ,
\label{criterion2}
\end{equation}
with $\Omega\equiv v_\phi /r$ and $I\equiv r B_\phi$.  

Eq.~(\ref{criterion2}) constitutes the second
important result of this work.  Once again, apart from the
new magnetoconvective term, eq.~(\ref{criterion2}) has the
same structure as the stability criterion derived for the much simpler
incompressible case (see eq.~[4.5] in Paper~I where the constant $\rho$ 
was normalized away). 

The new magnetoconvective term in eq.~(\ref{criterion2}) reveals a change in
coupling between the magnetic tension and the pressure differential
when $B_z$ is present (cf. eq.~[\ref{criterion1}] in \S~3 above): The magnetic-tension
force now works to oppose buoyant motions and the development of convection
in astrophysical fluids. This is because in the presence of gravity, 
pressure gradients are intrinsically negative but $B_\phi^2/r$ enters 
the second coefficient of the magnetoconvective term strictly with a positive sign.
The difference has its origin in the presence of the $B_z$ field that couples
fluid elements in the $z$ direction. When such elements attempt to move radially
in opposite directions, the magnetic tension acts as a restoring force and tries
to suppress these divergent motions. On the other hand, no such axial coupling
exists between elements in the $z$ direction in the case of only $B_\phi$ field, 
where the behavior of the magnetic tension is determined solely by the conservation
of specific azimuthal magnetic flux (see \S~3).

Finally we note that the magnetoconvective term does not vanish in homoentropic
fluids with $dK/dr$=0 because total pressure depends implicitly on the two magnetic-field
components. For the present case with $B_\phi$ and $B_z$ fields, using eq.~(\ref{pres_phi2}) 
and the conservation laws, 
the pressure differential of eq.~(\ref{ad2}) can be written in the equivalent form
\begin{equation}
\frac{dP}{dr} - \left.\frac{dP}{dr}\right|_{ad} = 
 \rho^\gamma \frac{dK}{dr} + \frac{1}{2}\left( \frac{1}{r^2}\frac{dI^2}{dr}
+ \rho^2 \frac{dH^2}{dr}\right)\ .
\label{dpk2}
\end{equation}
This form is used in the Appendix where we combine eqs.~(\ref{criterion2})
and (\ref{dpk2}) with $dK/dr$=0 and $I=0$, $B_\phi=0$ to confirm Newcomb's (1962) 
homoentropic stability criterion.

\section{Static Magnetized Nonhomoentropic Atmospheres}

We are now in a position to discuss clearly the stability of static
plane--parallel atmospheres with and without magnetic fields.  This is
a topic of considerable contemporary interest due to its astrophysical
applications to solar/stellar convection zones, stellar atmospheres,
and sunspots (e.g., Thomas \& Weiss 1992); and to accretion--disk magnetic
fields (e.g., Frank, King, \& Raine 1992; Duschl et al. 1994). Our results
for compressible flows can be easily reduced to this special case by
ignoring curvature--dependent terms (i.e., the centrifugal and the
magnetic--tension forces); and by replacing the derivatives with
respect to $r$ with derivatives with respect to $z$, where Cartesian
coordinates ($x, y, z$) are adopted throughout this section.  In this
new coordinate system, $z$ represents the vertical height in a
plane--parallel atmosphere or convection zone while the gravitational
force $g_z\equiv - {d\Phi_{\rm grav}}/{d z}$ points ``downward''
toward the negative $z$--axis.

In the simplest case of a static,
{\em unmagnetized}, plane--parallel atmosphere, the condition for
stability is given by the convective term in eq.~(\ref{criterion1})
or in eq.~(\ref{criterion2}) with $B_\phi = B_z = 0$, i.e., 
for stability to purely hydrodynamic convection
\begin{equation}
-\frac{dP_{fl}}{d z}\left(\frac{d P_{fl}}{d z}
-\left.\frac{d P_{fl}}{d z}\right|_{ad}\right) \geq 0\ ,
\label{convective}
\end{equation}
where the equilibrium density gradient is given by the equation
$dP_{fl}/dz = \rho g_z$. 
We note that the pressure gradient $dP_{fl}/dz < 0$ since $g_z < 0$, 
and eq.~(\ref{convective}) then leads to the Schwarzschild (1906)
criterion (cf. Clayton 1983)
\begin{equation}
\frac{dP_{fl}}{dz} \geq \frac{\gamma P_{fl}}{\rho}\frac{d\rho}{dz}\ ,
\label{clayton}
\end{equation}
or equivalently
\begin{equation}
\left|\frac{dP_{fl}}{dz}\right| \leq \left|\frac{dP_{fl}}{dz}\right|_{ad}\ .
\label{clayton2}
\end{equation}

The above analysis can be generalized now to include the
influence of a horizontal magnetic field ${\bf B} = B (z) {\bf e_x}$
in the direction of the unit vector ${\bf e_x}$. In order to apply the
interchange method correctly, we have to consider separately two
types of {\em vertical} perturbations in an atmosphere: those with
wavevectors ${\bf k}$ perpendicular to ${\bf B}$ (\S~5.1) and those
with ${\bf k}$ parallel to ${\bf B}$ (\S~5.2).

\subsection{Pure--Interchange Instability: ${\bf k\perp B}$}

We consider first the so--called ``pure interchange" case in which
the wavevector ${\bf k}\perp {\bf B}$ (Tserkovnikov 1960; Newcomb
1961). In this limit, our analysis of \S~3 is directly applicable.
Replacing $r$ by $z$ and $B_\phi$ by $B_x$, defining the conserved
specific flux by $\Phi_x\equiv B_x/\rho$, and ignoring 
curvature--dependent terms
in the equations of \S~3, the stability criterion becomes
(cf. eq.~[\ref{criterion1}])
\begin{equation}
-\frac{dP}{dz}\cdot\left(\frac{dP}{dz}-\left.\frac{dP}{dz}
\right|_{ad}\right)\geq 0\ ,
\label{Ts}
\end{equation}
where $P=P_{fl}+\rho^2 \Phi_x^2/2$ and
\begin{equation}
\left.\frac{dP}{dz}\right|_{ad}\equiv c_x^2\frac{d\rho}{dz}\ ,
\label{Paddef3}
\end{equation}
with
\begin{equation}
c_x^2 \equiv \frac{\gamma P_{fl} + B_x^2}{\rho} \ .
\label{cx}
\end{equation}
For a nonhomoentropic atmosphere with $g_z<0$, then $dP/dz < 0$ 
and this result reduces to the stability criterion
\begin{equation}
\left|\frac{d P}{d z}\right|\leq\left|\frac{d P}{d z}\right|_{ad}\ ;
\label{static0}
\end{equation}
and using
eqs.~(\ref{Paddef3}) and (\ref{cx}) along with the equilibrium expression $dP/dz=\rho
g_z$, we obtain the equivalent form (Tserkovnikov 1960)
\begin{equation}
\left|\frac{d \rho}{d z}\right|\geq\frac{\rho^2\left|g_z\right|}
{\gamma P_{fl}+B_x^2}\ .
\label{static1}
\end{equation}
This criterion for stability to pure interchanges is the direct
generalization of the purely hydrodynamic Schwarzschild criterion
(eq.~[\ref{clayton}] or eq.~[\ref{clayton2}]) and reduces to it in the
limit of $B_x\to 0$. 

Tserkovnikov's (1960) result for ${\bf k}\perp {\bf B}$ 
appears to suggest that the additional introduction
of a nonzero perturbation with ${\bf k}\parallel {\bf B}$ will not
modify the sufficiency of this criterion since the restoring forces
from field--line bending are naively expected to make a stabilizing
contribution. This idea was sharply criticized by Newcomb (1961) who
considered explicitly perturbations with ${\bf k}\parallel {\bf B}$
and obtained a different stability criterion which does not reduce to
Tserkovnikov's result in the limit of $k\rightarrow 0$.

This disagreement is simply a manifestation of {\em structural
instability} between the two models. As was explained in Paper~I,
structural instability is caused by any physical process that has the
power to destroy/replace/modify one or more conservation laws. 
We show this in \S~5.2 below, where we use the interchange
method to examine the instability caused by perturbations with ${\bf
k}\parallel {\bf B}$.

\subsection{Parker Instability: ${\bf k\parallel B}$}

We consider now perturbations in the so--called ``quasi--interchange''
case (Newcomb 1961).  The interchange involves two horizontal thin
finite ``strands" of fluid located initially along the magnetic field
at heights $z_1$ and $z_2 > z_1$.  We work in the limit of $k\to 0$;
hence, the derived criterion will only be sufficient for stability
(see related discussion in \S~4).

The quasi--interchange case is more subtle than the pure--interchange
case and its analysis proceeds along lines analogous to those
discussed in \S~4 above and in \S~4.1 of Paper~I: We allow for $k\neq
0$ perturbations by assuming that the fluid and the magnetic field
will remain unperturbed at $x=\pm\infty$. The physical interpretation
of this assumption is the following: During the displacement of fluid
element $i$ from height $z_i$ to height $z_j$ (where $i,j=1,2$), 
its mass is not conserved because it can flow along the distorted
field lines (i.e., in the direction of ${\bf k}$) toward or away from
the anchored (unperturbed) parts of the strand at $x=\pm\infty$.
Thus, just as in the $B_z$--field cases of \S~4 above and in Paper~I,
the ``horizontal infinity $x=\pm\infty$" serves in the present
model as a ``reservoir" through which mass is exchanged between
perturbed fluid elements.

Under the assumption that the boundary conditions at $x=\pm\infty$ do
not change, the quantity that dictates a conservation law is now 
the Bernoulli integral in the direction of ${\bf k}$ 
(as found from the poloidal projection of the Euler equation; 
see, e.g., Lovelace et al. 1986):
\begin{equation}
J\equiv\int\left.\frac{dP_{fl}}{\rho}\right|_{K={\rm const}}
 + \Phi_{\rm grav} = \frac{\gamma}{\gamma-1}K\rho^{\gamma-1} + 
\Phi_{\rm grav}\ .
\label{Be2}
\end{equation}

The above description is the essence of the so--called Parker (1966,
1975, 1979) instability (see also Shibata et al. 1989) for which we
are now equipped to apply the interchange method. The
relevant adiabatic density gradient in the quasi--interchange case is
determined from eq.~(\ref{Be2}) and the new conservation law $dJ/dz=0$,
viz.
\begin{equation}
\left.\frac{d\rho}{dz}\right|_{ad}\equiv\frac{\rho^2g_z}{\gamma P_{fl}}\ .
\label{rhoad}
\end{equation}

The magnetic field does not enter the Bernoulli integral because 
${\bf k\parallel B}$ perturbations
evolve only along field lines where the magnetic field does not exert tension. 
This again hints that we should interchange two fluid elements of different
masses $m_i\neq m_j$ and equal volumes $V_i = V_j$. By considering the
interchange of the contents of two equal volumes, the conservation of
specific magnetic flux is expressed by the requirement that the field strength
remain the same per unit mass; and, as was shown in \S~2, the conservation of the total 
mass contained in the two volumes is guaranteed.

Ignoring then curvature--dependent terms and keeping terms up to second order 
in the calculation of the gravity integrals of eq.~(\ref{DE2}), 
we obtain from eq.~(\ref{condition}) the condition
\begin{eqnarray}
\frac{\Delta E}{V} = & - & \frac{1}{2} \left[ 
 g_{z1} \rho_1 + g_{z2} (\rho_1 +\Delta\rho_{ad}) 
\right]\left(+\Delta z\right) \nonumber\\
& - & \frac{1}{2} \left[ 
 g_{z2} \rho_2 + g_{z1} (\rho_2 -\Delta\rho_{ad})
\right]\left(-\Delta z\right) \geq 0 ,
\label{big3}
\end{eqnarray}
or
\begin{equation}
\frac{\Delta E}{V} = \Delta z \left[g_z (\Delta\rho-\Delta\rho_{ad})\right]
\geq 0\ ,
\label{newDE}
\end{equation}
where $\Delta z = z_2 - z_1$ and $g_z = (g_{z1} + g_{z2})/2$. Since $\Delta z > 0$,
eq.~(\ref{newDE}) is equivalent to the sufficient criterion for stability 
\begin{equation}
g_z \left(\frac{d\rho}{dz}-\left. \frac{d\rho}{dz}\right|_{ad}
\right) \geq 0\ .
\label{sst}
\end{equation}
Recalling that $g_z < 0$ and using eq.~(\ref{rhoad}),
the stability criterion can be written in the general form (Newcomb 1961)
\begin{equation}
\left| \frac{d\rho}{dz}\right|\geq
\frac{\rho^2|g_z|}{\gamma P_{fl}}\ .
\label{static2}
\end{equation}
This inequality expresses the sufficient condition for suppression of
the Parker instability in the limit of $k\to 0$ perturbations. The
stability criterion does not depend on the magnetic field and is thus
different from eq.~(\ref{static1}). Thus, it is now clear from our 
analysis that
structural instability (the change in the conservation laws) explains
the difference of results in the cases with ${\bf k\perp B}$ and ${\bf
k\parallel B}$.

\section{Summary and Discussion}

We have derived general forms of the axisymmetric stability criteria
in compressible, rotating, magnetized, nonhomoentropic Couette flows
using an energy variational principle, the interchange or
Chandrasekhar's (1960) method. The results given in \S~3 and \S~4
generalize previous limited expressions obtained for incompressible
flows (Chandrasekhar 1981; Paper~I), for magnetostatic models
(Tserkovnikov 1960), and for suppression of magnetorotational 
instabilities in compressible homoentropic flows (Newcomb 1962;
Rogers \& Sonnerup 1986). 
To demonstrate the generality of our results and to assert the power 
of the interchange method, all these previous results are rederived as 
special cases of our two general stability criteria (eqs.~[\ref{criterion1}] 
and~[\ref{criterion2}]) 
and are placed in context in an Appendix to the paper.

We should point out that our resuls have been obtained by considering
adiabatic perturbations (thus also frozen-in magnetic fields). Our strongest 
and most general stability 
criterion~(\ref{criterion2}) differs from those obtained 
in magnetized models with finite conductivity and equal initial 
temperature along field lines, such as the criteria of \citet{b}.
In our treatment in \S~4, perturbations are constrained to conserve
angular velocity, axial current, and specific entropy; it is precisely
the gradients of these physical quantities that appear in the individual
terms that make up the stability criterion~(\ref{criterion2}). On the
other hand, the linear analysis of \citet{b} is based on isothermality
along field lines and the Boussinesq approximation, conditions that
impose the conservation of temperature between equally hot elements
coupled by the same field line. As a result then, temperature 
replaces specific entropy in
that stability criterion. A potential problem in this approach
is that coupled perturbed fluid elements that are maintained 
at the same temperature by the new conservation law are,
at the same time, not allowed by the Boussinesq 
assumption to change their pressures. Nonetheless, the structural 
instability discussed at length here and in Paper~I is
ever present in this model too, owing again to the change in
conservation laws effected by the axial field; as Balbus 
also points out, the hydrodynamic limit (the H{\o}iland criteria; 
see Tassoul 1978) can never be recovered even if 
the magnetic field is allowed to vanish.

We believe that our results are relevant in several areas of contemporary
astrophysical research. We outline below three areas along with 
brief discussions of the pertinent issues:

{\it (a) Magnetized stellar convection zones:}
The interchange method is directly applicable to static, magnetized,
plane--parallel atmospheres and stellar convection zones (see \S~5),
where it recovers well--known criteria for suppression of convective
motions and magnetic buoyancy (Schwarzschild 1906; Tserkovnikov 1960;
Newcomb 1961; Parker 1966, 1975, 1979). In addition, this method captures 
easily and explains clearly the difference between previously obtained 
results (Tserkovnikov 1960; Newcomb 1961) by identifying the apparent 
disagreement between magnetoconvective stability criteria 
(eq.~[\ref{static1}] and eq.~[\ref{static2}])
with a typical case of structural instability.

{\it (b) Magnetized low-luminosity accretion flows:}
We believe that our work also sheds some light to 
the recent controversy concerning the radiatively inefficient accretion 
flows and the direction of angular momentum transport in associated models
\citep{nia, nqia, bh3}. Our analysis of \S~4 delineates the 
role of rotation, magnetic fields, and compressibility
to the stability of such flows. As emphasized here and in Paper~I, 
flow stability as well as angular momentum and magnetic flux transport 
are governed by the demand that certain 
conservation laws be respected by perturbed fluid elements. 
Specifically, for elements coupled by the same poloidal field line:

\begin{enumerate}
\item The constant $\Omega = L/r^2$ law leads to outward angular momentum transport---as some fluid elements move outward their $r$ increases, 
thus their $L$ must also increase in order to maintain the same $\Omega$.

\item At the same time, the constant $I = r B_\phi$ law leads to inward 
azimuthal magnetic flux transport---as some fluid elements move inward 
their $r$ decreases, thus their $B_\phi$ must increase in order to 
maintain the same $I$.
\end{enumerate}

The sense of transport of these quantities is absolutely fixed by the mere presence
of an axial magnetic field component, the component that is also directly
responsible for the existence of magnetorotational instability.
The magnetized flow model of \S~4 makes it quite clear that
the non-magnetic stability criteria and the sense of convective transport 
of associated quantities cannot be smoothly recovered,
even in the limit of $B_z \to 0$. It is thus hard to see how angular 
momentum could be transported inward in the presence of a frozen-in axial field. 
This however might be the case if the field could ``slip" through the plasma 
(in which case the ideal MHD conditions do not apply), a situation that would 
immediately also invalidate the conservation laws~(21) and~(22). 

In relation to the role of the $B_{\phi}$ component, we re-iterate here 
that, in the presence of an axial field, this component can reduce significantly 
the effects of bouyancy, as shown by eq.~(\ref{criterion2}) above.
It is unfortunate that the $B_\phi$ component has been ignored in many previous 
studies with no good justification, other than that
its absence makes the linear--stability
analyses manageable. We believe that such simplification of the
dynamical equations is not justified in accretion disks for the following reasons:
(i)~Differential rotation will twist around any initial weak radial field
and will continue to amplify the resulting toroidal field. (ii)~The toroidal
field produced in this way will be transported inward, just as the
angular momentum is transported outward. It is therefore
certain that azimuthal magnetic flux will pile up in the inner regions
of magnetized disks where it will dominate the dynamics of these regions.
We discuss this point in more detail in part (c) below.

While the instability criteria that we have derived are effectively local, it is
interesting to speculate about possible applications of our
results to global flows, in particular those of ADAF--type, in which
all sound speeds scale as $c_s \propto r^{-1/2}$ and the temperature scales as
$T \propto r^{-1}$ (Ichimaru 1977; Narayan \& Yi 1994). 
A similar issue was raised in Paper~I, where
we speculated about whether nonlinear simulations might lead to 
flows with constant $\Omega$ and $B_{\phi} \propto 1/r$ that
hover around the incompressible $\Delta E \simeq 0$ stability limit. 
Now, it is clear from the outset that $\Omega$ cannot be constant 
in an ADAF. However, the formulation of our problem 
still allows us to examine the possibility of having a flow with $\Delta E 
\simeq 0$ on {\em all scales}, i.e. a flow that
operates at the edge of instability at all relevant scales.
Then the condition $\Delta E \simeq 0$ requires that all terms 
in eqs.~(\ref{criterion1}), (\ref{criterion2}) be of the same order 
of magnitude; replacing the $d/dr$ derivatives by $1/r$ in either equation,
we find that 
\begin{equation}
\Omega^2 \simeq \frac{B^2_{\phi}}{\rho r^2} \ ,
\label{equip}
\end{equation}
over all scales,
i.e., that the magnetic field should be in rough equipartition with the 
rotational kinetic energy! This condition is independent of the specific radial
scalings obeyed by $\Omega (r)$, $B_{\phi} (r)$, and $\rho (r)$.

It is interesting to note that the global condition~(\ref{equip}) obtains only in the compressible case, since in an incompressible flow the density drops out of the stability criterion (cf. eq.~[4.6] of Paper I). It is 
also interesting that such equipartition was actually 
assumed in the original ADAF formulation for the explicit purpose
of reducing the adiabatic index of the flow to below $5/3$.  
Finally, we point out that although eq.~(\ref{equip})
does not uniquely determine the 
scalings of $\Omega (r)$, $B_{\phi} (r)$, and $\rho (r)$ 
of the flow, it still imposes an important constraint on the
strength of the toroidal field for a given accretion geometry.

{\it (c) Magnetized jet-like outflows:}
Our results are useful to the analysis of the vertical structure
of fully ionized magnetized accretion disks.  The stability criterion
(\ref{static1}) may resolve the issue of how much azimuthal magnetic
flux can be stored in the interior of an accretion disk before it
begins to escape buoyantly to the surface and, from there on, to the
surrounding region.  The condition for stability against buoyant
escape of a whole azimuthal magnetic--flux tube (eq.~[\ref{static1}]) is
{\em weaker} than the condition against the ``nonaxisymmetric" Parker
instability (eq.~[\ref{static2}]) which requires steeper density
gradients to stabilize ${\bf k}\parallel {\bf B}$ perturbations. 
However, if we consider axisymmetric
perturbations in an axisymmetric accretion disk with only a
toroidal--field component, then only eq.~(\ref{static1}) is relevant
and the criterion for stability against vertical magnetic buoyancy in
an homoentropic disk can be written in the form (cf.  eq.~[\ref{dpk}]
with $dK/dr = 0$ and without curvature--dependent terms; and eq.~[\ref{Ts}] with
$dP /dz < 0$)
\begin{equation}
\frac{d\Phi^2}{dz}\equiv \frac{d}{dz}\left(
\frac{B_\phi^2}{\rho^2}\right) \geq 0\ .
\label{phix}
\end{equation}
A similar condition for suppression of radial magnetoconvective
motions, namely 
\begin{equation}
\frac{dH^2}{dr}\equiv \frac{d}{dr}\left(
\frac{B_z^2}{\rho^2}\right) \geq 0\ ,
\label{phiy}
\end{equation}
is obtained from eqs.~(\ref{criterion2}) and~(\ref{dpk2}) in the
special case of a nonrotating ($\Omega = 0$) homoentropic ($dK/dr = 0$)
disk with a dynamically important, purely axial ($I = 0$, $B_\phi = 0$)
magnetic field $B_z$, and with $dP/dr < 0$ (see also Anzer [1969] and 
Lubow \& Spruit [1995]). Both conditions are expressions of the
Kruskal--Schwarschild (1954) criterion for stability of an atmosphere
supported against gravity only by fluid--pressure and ``horizontal" 
magnetic--pressure gradients.

We heretofore focus on the significance of eq.~(\ref{phix}) for jets
and accretion disks in active galactic nuclei.  It is generaly
believed that the toroidal magnetic field, which is naturally produced
by differential winding of any weak radial magnetic--field component,
is always limited by buoyant escape to remain much below equipartition
levels (see, e.g., D'Silva \& Chakrabarti 1994). On the other hand,
numerical simulations (Shibata, Tajima, \& Matsumoto 1990)
indicate that under some (unspecified) conditions, buoyant escape
appears to be suppressed and does not prevent the toroidal magnetic
field from growing to equipartition values (see also Galeev, Rosner,
\& Vaiana 1979).  In the same context, Contopoulos (1995) has
considered the role of the toroidal magnetic field in generating
collimated energetic bipolar jet outflows in accretion disks.  The
issue of the initial buildup of the toroidal field was however left as
an open question.

We now suggest the following scenario: Even if the field is initially
confined to the interior of the disk, as inflow of matter and
differential winding of the radial magnetic field continue,
inequality~(\ref{phix}) can be reversed.  Then, magnetoconvective
instability will set in and azimuthal magnetic flux will leak out of
the disk, in an attempt to return to a stable vertical stratification
with $d\Phi^2/dz\geq 0$.  Now, one has to consider the relevant
timescale for radial inflow of matter and azimuthal magnetic flux: 
If the growth of $\Phi$ in the disk's interior proceeds
dynamically (because of an implosive magnetorotational instability),
then the buoyant escape will take place explosively---just as during
the release phase of a typical laboratory plasma gun---and will
presumably result in the discharge of highly energetic, axially moving
``blobs'' of disk material. This mechanism may account for the
relativistically moving blobs observed by VLBI in jet--like outflows
in active galactic nuclei (e.g., Hughes 1991; Burgarella, Livio, \&
O'Dea 1994).

Condition~(\ref{phix}) for suppression of vertical magnetoconvective
motions is everywhere satisfied if the fluid density decreases
sufficiently fast (e.g. exponentially) with height from the mid--plane
of an accretion disk. In contrast, if the density does not decrease
sufficiently fast with height, then inequality~(\ref{phix}) may be
permanently reversed at some height. Such a reversal will lead to slow
leakage of the inflowing azimuthal magnetic flux. 
This continuous leakage in the $z$ direction may
prevent altogether the dynamical buildup of the azimuthal flux which
is expected to produce by discharge a relativistic jet outflow (see
the discussion above and in Contopoulos 1995).  This tentative
conclusion deserves further investigation in relation to the dichotomy
between radio--loud and radio--quiet quasars.



\acknowledgments

This work was supported in part by a Chandra Guest Observer grant.



\appendix

\section{Special Forms of the Stability Criteria}

\subsection{Comparisons with the Results of Tserkovnikov (1960)}

\noindent
(a)~In order to compare our stability criterion for an homoentropic
magnetostatic cylindrical equilibrium threaded by a purely toroidal
magnetic field to the corresponding expression found by Tserkovnikov (1960) 
and confirmed by Newcomb (1962), we combine: 
\begin{enumerate}
\item Eq.~(\ref{eq}) with $v_\phi$=0, $g_r$=0, and $B_z$=0;
\item Eq.~(\ref{criterion1}) with $L$=0; 
\item Eq.~(\ref{dpk}) with $dK/dr$=0;
\end{enumerate}
and we obtain the form
$$
\frac{d\Phi^2}{dr}\leq 0\ .\eqno(A1)
$$
After straightforward algebraic reductions, this stability criterion
can be cast in the equivalent form
$$
\frac{d\ln{B_\phi}}{d\ln{r}}\leq\frac{\gamma P_{fl}-B_\phi^2}
{\gamma P_{fl}+B_\phi^2}\ ,\eqno(A2)
$$
that was derived by Tserkovnikov (1960). 

We see now how the complexity in Tserkovnikov's result obscures the basic physics 
of the magnetostatic problem. Eq.~(A2) is just a complicated relation between
$B_\phi$ and $P_{fl}$ with no obvious physical meaning. On the other hand, the
equivalent form eq.~(A1) reveals that, despite the homoentropic assumption, unstable 
convective modes 
can still exist if the magnetic field establishes a superadiabatic pressure gradient.
This result is valid in the incompressible case as well (Chandrasekhar 1981; eq.~[3.7]
in Paper~I). Its general applicability should have been expected because the
second-order pressure differential in eq.~(\ref{dpk}) depends also on
the magnetic flux gradient (and not on the entropy gradient alone). 

\noindent
(b)~Along similar lines, it is expected that the stability criterion for a magnetostatic
equilibrium threaded by both toroidal and axial field is expressed by a simple condition 
involving the radial variation of the axial current. This result has been previously 
derived in the incompressible case (Chandrasekhar 1981; 
eq.~[4.7] in Paper~I) but it has not been
generalized for compressible models until now. We combine: 
\begin{enumerate}
\item Eq.~(\ref{eq}) with $v_\phi$=0 and $g_r$=0;
\item Eq.~(\ref{criterion2}) with $\Omega$=0;
\end{enumerate}
and we obtain the form
$$
\frac{dI^2}{dr}\leq 0\ .\eqno(A3)
$$
Note that eq.~(\ref{dpk2}) plays no role here in interpreting 
the result because the reductions
$v_\phi = 0$ and $g_r = 0$ in eq.~(\ref{eq}) imply that 
$dP/dr + B_\phi^2 /r \equiv 0$ and the last
term of eq.~(\ref{criterion2}) is then eliminated from the stability criterion.
Therefore eq.~(A3) is generally valid for any distribution of field $B_z (r)$ 
and axial flux $H (r)=B_z(r)/\rho (r)$ in magnetostatic models.

\subsection{Comparisons with the Results of Newcomb (1962)}

The stability criteria obtained by Newcomb (1962) were also discussed 
in the Appendix of Paper~I where we reported erroneously that Newcomb's 
derivations miss an additional magnetoconvective term. This statement
is not correct---it turns out that the slightest error in the application
of the interchange method generates additional spurious terms. However,
the mathematical reductions reported in the Appendix of Paper~I were
performed only on Newcomb's equations and they are correct.

Using the general stability criteria reported in this paper, 
we can reproduce now Newcomb's results precisely: 

\noindent
(a)~In order to compare our stability criterion for an homoentropic fluid
threaded by a purely toroidal magnetic field to the corresponding
expression obtained by Newcomb (1962), we combine: 
\begin{enumerate}
\item Eq.~(\ref{eq}) with $g_r$=0 and $B_z$=0;
\item Eq.~(\ref{criterion1}) with $\Phi = B_\phi /(\rho r)$; 
\item Eq.~(\ref{dpk}) with $dK/dr$=0 and $\Phi = B_\phi /(\rho r)$;
\end{enumerate}
and we obtain the form
$$
\frac{1}{r^3}\frac{d}{d r}\left(\rho L^2\right)
-\frac{2}{r}B_\phi\frac{d B_\phi}{d r}
+\frac{2B_\phi^2}{r^2} \geq \frac{1}{\rho c_\phi^2}
\left(\frac{\rho L^2 - 2 r^2 B_\phi^2}{r^3}\right)^2 \ ,\eqno(A4)
$$
which is identical to eq.~(8.11) in Newcomb (1962).
We see again how the complexity in Newcomb's result obscures the basic physics 
of the problem. Eq.~(A4) is just a complicated relation between all the
relevant quantities with no obvious physical meaning. On the other hand, the
equivalent form of the stability criterion~(\ref{criterion1}) 
along with the definition~(\ref{dpk}) can be physically interpreted with
ease as the interplay between gradients in the three conserved
quantities (specific angular momentum, azimuthal flux, and entropy), 
as was done in \S~3 above.

\noindent
(b)~In order to compare our stability criterion for an homoentropic fluid
threaded by a purely axial weak magnetic field to the corresponding
expression obtained by Newcomb (1962), we combine:
\begin{enumerate}
\item Eq.~(\ref{eq}) with $g_r$=0 and $B_\phi$=0;
\item Eq.~(\ref{criterion2}) with $I$=0 and $B_\phi$=0; 
\item Eq.~(\ref{dpk2}) with $dK/dr$=0, $I$=0, and $H = B_z /\rho$;
\end{enumerate}
and we obtain the form
$$
\frac{d}{d r}\left(\rho L^2\right) \geq \frac{4\rho L^2}{r} +
\frac{r^3}{\rho c_z^2}\left(\frac{\rho L^2}{r^3}\right)^2\ ,\eqno(A5)
$$
which is identical to eq.~(10.4) in Newcomb (1962) if we let
$B_z\to 0$ and $c_z\to c_o$ (see eq.~[\ref{c_phi2}]).
Once again, eq.~(A5) is just a complicated relation between all the
relevant quantities with no obvious physical meaning. On the other hand, the
equivalent form of the stability criterion~(\ref{criterion2}) 
along with the definition~(\ref{dpk2}) can be physically interpreted with
ease as the interplay between gradients in the three conserved
quantities (angular velocity, axial current, and specific entropy), 
as was done in \S~4 above.

\subsection{Comparisons with the Results of \citet{rs}}

\citet{rs} have obtained a form of the stability criterion
in the case of an homoentropic fluid threaded by toroidal field only, 
and they
pointed out that their result agrees with the condition found by
Newcomb (1962), eq.~(A4) above. Although technically correct, their
result is subject to two fundamental assumptions: (a)~The fluid is uniformly
rotating, while Newcomb's result is valid for a differentially rotating fluid.
(b)~The fluid is gravitating, while Newcomb's result was derived for a
nongravitating fluid. Therefore, according to (a), the criterion
of Rogers \& Sonnerup is only a special case of Newcomb's result, 
but according to (b), it is more general than Newcomb's result. Obviously,
we must exercise caution when we reduce our result to the form of \citet{rs},
and we should not use eq.~(A4) because it was derived under the assumption 
that $g_r$=0.   

In order to compare our stability criterion for an homoentropic, 
uniformly rotating, gravitating fluid
threaded by a purely toroidal magnetic field to the corresponding
expression obtained by \citet{rs}, we combine: 
\begin{enumerate}
\item The assumption that $\Omega = {\rm const}$;
\item Eq.~(\ref{eq}) with $v_\phi = \Omega r$, $B_z$=0, and $g_r = -d\Phi_{grav}/dr$;
\item Eq.~(\ref{criterion1}) with $L=\Omega r^2$ and $\Phi = B_\phi /(\rho r)$; 
\item Eq.~(\ref{dpk}) with $dK/dr$=0 and $\Phi = B_\phi /(\rho r)$;
\item The definitions of \citet{rs}: $P_{fl}\equiv K\rho^\gamma$,
$P_{mag}\equiv B_\phi^2 /2$, $P\equiv P_{fl} + P_{mag}$, 
and $\Gamma\equiv\Phi_{grav} - \Omega^2 r^2 /2$;
\end{enumerate}
and we obtain the form
$$
\frac{2}{r}\left(\frac{2 P_{mag}}{r} - \frac{dP_{mag}}{dr}\right)
- \frac{1}{\gamma P_{fl} + 2P_{mag}}\left(\frac{2 P_{mag}}{r} - \frac{dP}{dr}\right)^2
- \frac{d\rho}{dr}\frac{d\Gamma}{dr} + 4 \rho \Omega^2 \geq 0\ ,\eqno(A6)
$$
which is identical to eq.~(15) in \citet{rs}. We note, once again, that 
$\Omega = {\rm const}$ in this derivation, hence 
$d\Gamma /dr \equiv - \left(g_r + \Omega^2 r\right)$ in eq.~(A6).

In this case too, the above result contains substantially less physics than our
stability criterion~(\ref{criterion1})---no gradients in the angular velocity
or in the specific entropy---yet it is hard to deduce the elegant
physical interpretation afforded by our equivalent form (eqs.~[\ref{criterion1}]
and~[\ref{dpk}] with $\Omega = {\rm const}$ and $dK/dr$=0): 
$$
4\Omega^2 - \rho r \frac{d\Phi^2}{dr}
- \frac{r^2}{2 c_\phi^2}\left(\frac{dP}{dr} - \frac{B_\phi^2}{r}\right) 
\frac{d\Phi^2}{dr} \geq 0\ .\eqno(A7)
$$
Here the first term is the epicyclic frequency due to solid--body rotation 
and has a stabilizing influence. The second term depends on the azimuthal 
flux gradient. The last term is convective and is amplified by the toroidal
field, but its sign depends only on the sign of the gradient $d\Phi^2 /dr$.
This is because $dP/dr < 0$ and thus the coefficient $( dP/dr - B_\phi^2 /r )$ 
is always negative in cases of astrophysical interest.
Consequently, the pressure differential is determined to be superadiabatic
or subadiabatic solely by the radial distribution of $\Phi^2$, given that the
fluid is homoentropic (eq.[\ref{dpk}] with $dK/dr=0$).





\end{document}